\documentclass[aps,prl,twocolumn,showpacs,superscriptaddress]{revtex4}

\usepackage{graphicx}
\usepackage{dcolumn}
\usepackage{bm}

\def\vec#1{\mbox{\boldmath $#1$}}

\def\vec#1{\mbox{\boldmath $#1$}}
\def\dfrac#1#2{{\displaystyle\frac{#1}{#2}}}
\renewcommand{\epsilon}{\varepsilon}

\def\Gammai{\mbox{$\mit \Gamma$}}

\begin{document}

\title{Holographic Storage of Multiple Coherence Gratings in a Bose-Einstein Condensate}

\author{Yutaka Yoshikawa}
\email{yutaka@phys.c.u-tokyo.ac.jp}
\affiliation{Institute of Physics, University of Tokyo, 3-8-1, Meguro-ku, Komaba, Tokyo 153-8902, Japan.}
\affiliation{PRESTO, CREST, Japan Science and Technology Agency, 4-1-8 Honcho,
Kawaguchi, Saitama, Japan.}
\author{Kazuyuki Nakayama}
\affiliation{PRESTO, CREST, Japan Science and Technology Agency, 4-1-8 Honcho,
Kawaguchi, Saitama, Japan.}
\author{Yoshio Torii}
\affiliation{Institute of Physics, University of Tokyo, 3-8-1, Meguro-ku, Komaba, Tokyo 153-8902, Japan.}
\affiliation{PRESTO, CREST, Japan Science and Technology Agency, 4-1-8 Honcho,
Kawaguchi, Saitama, Japan.}
\author{Takahiro Kuga}
\affiliation{Institute of Physics, University of Tokyo, 3-8-1, Meguro-ku, Komaba, Tokyo 153-8902, Japan.}
\affiliation{PRESTO, CREST, Japan Science and Technology Agency, 4-1-8 Honcho,
Kawaguchi, Saitama, Japan.}

\date{\today}

\begin{abstract}
We demonstrate superradiant conversion between a two-mode collective atomic state and a single-mode light field in an elongated cloud of Bose-condensed atoms. Two off-resonant write beams induce superradiant Raman scattering, producing two independent coherence gratings with different wave vectors in the cloud. By applying phase-matched read beams after a controllable delay, the gratings can be selectively converted into the light field also in a superradiant way. Due to the large optical density and the small velocity width of the condensate, a high conversion efficiency of $> 70$ \% and a long storage time of $> 120$ $\mu$s were achieved.
\end{abstract}


\maketitle

Atomic ensemble is a promising candidate for quantum memory, in which quantum information can be stored as a long-lived ground-state coherence. In contrast, photons are ideal for long-distance quantum-state transfer since they travel fast and are robust against perturbations from environment. The conversion of quantum states between atoms and photons has thus been an important subject in recent years  \cite{Duan1, Lukin, Chou1, Chaneliere, Black, van der Wal, Kuzmich, Balic, Matsukevich, Chou2, Duan2}. In particular, on the basis of spontaneous Raman scattering and coherent reverse processes \cite{Duan1, Lukin}, various applications, including generation of single photons \cite{Chou1, Chaneliere, Black}, correlated photon pairs \cite{van der Wal, Kuzmich, Balic} and entangled states \cite{Matsukevich, Chou2}, have been successfully demonstrated.

In general, indistinguishable atoms interacting with a single-mode light field shows enhanced light scattering by a factor of $N$, where $N$ is the number of atoms. This cooperative effect is the key factor for the efficient conversion from an atomic coherence to the light field in a well-defined spatial mode \cite{Duan1, Duan2}. The figure of merit is given by the optical density of atoms along the mode axis $N \eta$, where $\eta \equiv \sigma_a / A$ is the single-atom optical density. Here $\sigma_a$ is the atomic absorption cross section and $A$ is the cross section of the interaction region perpendicular to the mode axis. In free space $\eta$ is scaled as the solid angle determined by the diffraction of scattered photons $\approx \lambda^2 / D^2$, with the light wavelength $\lambda$ and the cloud diameter $D$. Therefore, $N \eta$ roughly gives $R_{\rm m} / R$, the ratio of the photon scattering rate into the desired mode $R_{\rm m}$ to that into the other mode $R$. The conversion efficiency can then be written as $P = R_{\rm m}/(R_{\rm m} + R) \approx N \eta / (1 + N \eta)$ \cite{Black}.

To achieve large $N \eta$, the most popular method is to use forward Raman scattering from an elongated interaction volume, determined by the pump-beam path in the atomic ensemble \cite{Duan1, Chou1, Chaneliere, van der Wal, Kuzmich, Balic, Matsukevich, Chou2}. An alternative approach is to put the atoms into an optical cavity \cite{Black}, where the photon emission into the cavity mode is enhanced by the Purcell factor $\approx 2 \zeta N \eta$, where $\zeta$ is the number of round trips of photons in the cavity \cite{Horak}. The third approach is to use an elongated atomic cloud, in which the optical modes along the long axis of the cloud (referred to as ``end-fire modes'') have automatically large $N \eta$ \cite{Gross}. This system has been recently investigated in the context of superradiant light scattering from Bose-Einstein condensates (BECs) \cite{Inouye1, Schneble1, Yoshikawa1, Yoshikawa2}. Note that, in the second and the third approach, the spatial modes possessing large $N \eta$ are independent of the pump-beam direction, then allowing simultaneous writing of many coherence gratings in the single cloud. In addition, these gratings can be independently converted into photons because of the Bragg selectivity in the read process.

In this work, we experimentally demonstrate holographic storage of two coherence gratings in a single BEC cloud, from which a single-mode light field can be independently retrieved at arbitrary timing. The write-read process was performed with phase-matched anti-Stokes and Stokes superradiant Raman scattering (SRS) \cite{Schneble1, Yoshikawa1}. Although the previous demonstrations have been limited to the storage of one state per cloud \cite{Chou1, Chaneliere, Black, van der Wal, Kuzmich, Balic, Matsukevich, Chou2}, the present scheme would open the possibilities of realizing holographic quantum memory, in which many quantum states can be simultaneously stored in a single cloud.

\begin{figure}[tttt]
\begin{center}
    \scalebox{0.8}{\includegraphics{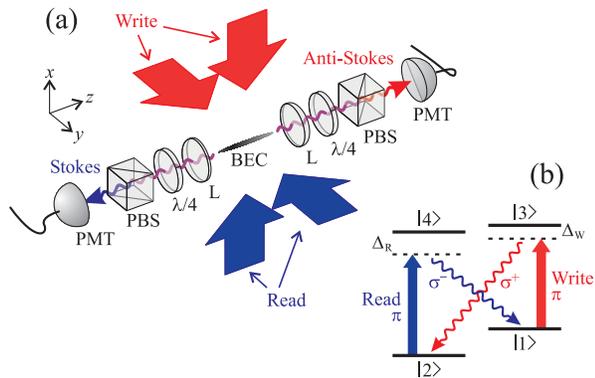}}
\end{center}
	\caption{(Color online). (a) Geometry and (b) energy-level diagram of the experiment. L: lens, $\lambda/4$: quarter-wave plate, PBS: polarization beam splitter, and PMT: photo-multiplier tube.}
	\label{Fig.1}
\end{figure}

Let us consider the situation that the BEC containing $N$ atoms scatters $n$ photons (assuming $n \ll N$) from an off-resonant ``write'' beam to the end-fire mode (see, also Fig.\ \ref{Fig.1}). Then, exactly $n$ atoms in the BEC receive the recoil momentum with a wave vector $\vec{q} = \vec{k}_{\rm W}-\vec{k}_{\rm E}$, where $\vec{k}_{\rm W}$ and $\vec{k}_{\rm E}$ is the wave vector of the write beam and the end-fire mode, respectively. As a result, the BEC and the $n$-recoiling atoms form a ``coherence grating'' $\propto \sqrt{N n} \exp(i \vec{q} \vec{r})$. Then, the information of the photon number and the phase is stored in the cloud in the form of this grating. To read out this information, another light beam (``read beam'') with a wave vector $\vec{k}_{\rm R}$ is applied to the atoms. If $\vec{k}_{\rm R} \approx -\vec{k}_{\rm W}$ to fulfill the phase-matching (Bragg) condition $\vec{k}_{\rm R} + \vec{k}_{\rm E} \approx -\vec{q}$, the read beam diffracts off the grating and the $n$-recoiling atoms are converted back into $n$ photons in the opposite end-fire mode \cite{van der Wal}. In contrast, if $\vec{k}_{\rm R} \neq -\vec{k}_{\rm W}$, the Bragg diffraction does not occur and the grating is kept stored in the cloud. This means that, if we apply many write beams with different wave vectors, many gratings are simultaneously stored in the cloud and each of them is read out only by the phase-matched read beam. A similar technique has been employed for holographic classical data storage \cite{Heanue}.

The scattering rate of the read photons is proportional to the square of the grating amplitude $N n$, which shows $N$-fold enhancement compared to the spontaneous scattering rate. We emphasize here that the atomic state with the grating is formally equivalent to the Dicke state \cite{Dicke} due to full symmetry against the exchange of arbitrary atoms. Therefore, the origin of this read-photon diffraction is physically the same as the collective enhancement in the Dicke state \cite{Ketterle}. This implies that BECs are not essential in the present scheme, although practical due to their large $N \eta$ and long coherence time \cite{Yoshikawa2}. Cold atoms in a cavity \cite{Black, Slama} and trapped Fermi gases \cite{Roati} might be other candidates as a storage medium.

Figure \ref{Fig.1} shows the geometry and the energy-level diagram of the experiment. A BEC of $^{87}$Rb atoms, just after the release from a magnetic trap \cite{footnote}, was illuminated by off-resonant light beams from radial direction. For the demonstration of multiplexed storage of the gratings, we prepared two pairs of counter-propagating write and read beam, with their optical axes being orthogonal to each other. The BEC contained $N = 8 \times 10^5$ atoms with the size of $l = 230$ $\mu$m in length and $d = 14$ $\mu$m in diameter. The Fresnel number of the cloud $F = \pi d^2/(4 \lambda l)$ was $0.85$ with $\lambda =$ 795 nm, so that the end-fire modes can be treated as a single transverse mode \cite{Gross}. For these modes, $N \eta \approx 1600$ and the calculated conversion efficiency was $> 99$\%.

The write and the read beams were $\pi$-polarized with respect to $z$ axis and tuned nearly to the $|1 \rangle \equiv |5{\rm S}_{1/2}; F = 2, m_F = 2 \rangle \to |3 \rangle \equiv |5{\rm P}_{1/2}; F = 2, m_F = 2 \rangle$ and the $|2 \rangle \equiv |5{\rm S}_{1/2}; F = 1, m_F = 1 \rangle \to |4 \rangle \equiv |5{\rm P}_{1/2}; F = 1, m_F = 1 \rangle$ transition, respectively. The detunings from the resonance were set to be $\Delta_{\rm W} = 2 \pi \times -1.0$ GHz for the write and $\Delta_{\rm R} = 2 \pi \times -1.8$ GHz for the read beam. The write beam induces anti-Stokes SRS \cite{Yoshikawa1}, by which the atoms in the initial state $|1 \rangle$ are pumped into the state $|2 \rangle$, accompanied by the emission of $\sigma^+$-polarized photons in the end-fire mode. The read beam induces Stokes SRS \cite{Schneble1}, whereby the recoiling atoms are pumped back to the state $|1 \rangle$ with the emission of $\sigma^-$-polarized photons along the opposite direction to the anti-Stokes photons. Notice that SRS occurs both along $\pm z$ directions, resulting in two pairs of correlated anti-Stokes and Stokes photons. However, we monitored only one pair by selecting the polarization of the photons with quarter-wave plates and polarization beam splitters in front of photo-multiplier tubes.

\begin{figure}[tttt]
\begin{center}
    \scalebox{0.8}{\includegraphics{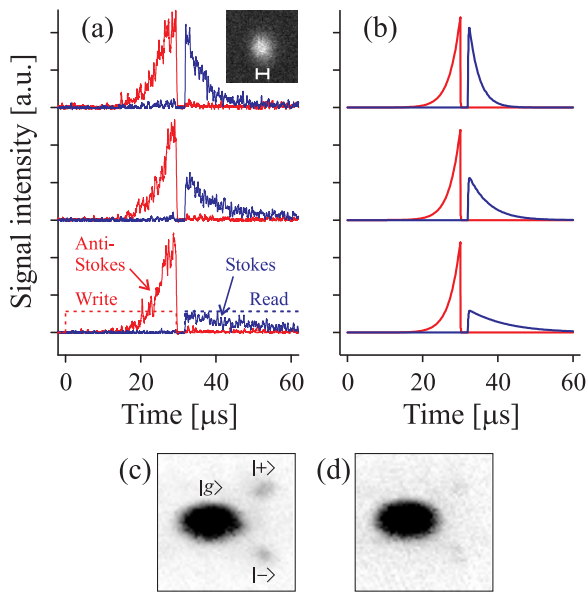}}
\end{center}
	\caption{(Color online). (a) Observed superradiant anti-Stokes and Stokes pulses. Timing of the light beams is shown as the dashed lines in the lowest trace. The intensity of the write beam was $I_{\rm W} = 41$ mW/cm$^2$ and that of the read beam was changed to $I_{\rm R} = 510$, $255$, and $128$ mW/cm$^2$ from upper to lower. The spatial profile of the end-fire mode is shown in the inset. A white bar represents the diffraction angle $\theta_{\rm d}$. (b) Numerical simulations based on Eqs. (\ref{inversion})-(\ref{coherence}). The parameters were $\Gammai_a = 2 \pi \times 1.3$ kHz and $\gamma_{\rm AS} = 2 \pi \times 160$ Hz. Three traces correspond to the scattering rate ratio $\gamma_{\rm S} / \gamma_{\rm AS} = 1.0$, $0.5$, and $0.25$ from upper to lower. (c) and (d) show the optical density of atoms after $34$-ms ballistic expansion for just after the illumination of a $30$-$\mu$s write pulse and a successive $30$-$\mu$s read pulse, respectively.}
\label{Fig.2}
\end{figure}

Here, we first present the properties of single-mode storage with one write-read beam pair. Figure \ref{Fig.2}(a) shows typical signals of the anti-Stokes and the Stokes pulses. The number of photons contained was $\sim 3 \times 10^4$. The inset shows the intensity distribution of the end-fire mode, directly taken by a charge-coupled device camera. The profile was found to be a Gaussian shape with a divergence angle consistent with the diffraction angle $\theta_{\rm d} = \lambda/d$ (depicted as a white bar). This feature is crucial for additional single-mode operations to the retrieved photons \cite{Duan1, Chaneliere, Kuzmich, Matsukevich, Chou2}.

The observed superradiant pulses show a tendency of exponential growth for the write and exponential decay for the read process. This behavior can be qualitatively interpreted as the inversion of cascaded transitions in the Dicke state \cite{Gross, Dicke}: In the dressed-state picture, the state of atoms coupled to the write beam corresponds to a highly excited Dicke state in a two-level system, leading to an exponential increase of the spontaneous emission rate into the bare state $|2 \rangle$. However, by switching the write beam off and the read beam on, the ladder of the dressed Dicke state is nonadiabatically inverted. As a result, the highly excited state changes to a low excited state, from which the spontaneous emission to the state $|1 \rangle$ gets exponentially weaker as it occurs.

Quantitatively, the dynamics of the SRS is described by the coupled equations for three atomic states $|g \rangle \equiv |1 \rangle \otimes |\vec{p} = \vec{0} \rangle$ and $|\pm \rangle \equiv |2 \rangle \otimes |\vec{p} = \hbar (\vec{k}_{\rm W} \pm \vec{k}_{\rm E}) \rangle$, where $\vec{p}$ is the momentum of atoms. Defining the single-atom density matrix element as $\rho_{ij}$ ($i, j = g, \pm$), the equations of motion for an inversion $U \equiv \rho_{++} + \rho_{--} - \rho_{gg}$ and a coherence $V_\pm \equiv 2\>{\rm Im} (\rho_{g \pm})$ are written as \cite{Inouye2},
\begin{eqnarray}
\dot U &=& G (\gamma_{\rm AS} - \gamma_{\rm S}) ( V_+ + V_- )^2 \label{inversion} \\
\dot V_\pm &=& \left[2 G (\gamma_{\rm S} - \gamma_{\rm AS}) U - \dfrac{\Gammai_a}{2} \right] V_\pm, \label{coherence}
\end{eqnarray}
where $\Gammai_a$ is the decoherence rate of atoms, $G = 3 N \lambda^2 /(8 \pi^2 d^2)$ is the superradiant gain coefficient related to the optical density $G = N \eta / 16$, and $\gamma_{\rm AS}$ ($\gamma_{\rm S}$) is the single-atom Raman scattering rate for the write (read) process. Here, we have omitted the effect of end-fire mode photons \cite{Schneble2} and the propagation effect \cite{Zobay} for simplicity. The calculated rate of the anti-Stokes SRS, $N G \gamma_{\rm AS} V_+^2$, and that of the Stokes SRS, $N G \gamma_{\rm S} V_+^2$, are plotted in Fig.\ \ref{Fig.2}(b). They show fairly well agreement with the experimental results.

\begin{figure}[tttt]
\begin{center}
    \scalebox{0.8}{\includegraphics{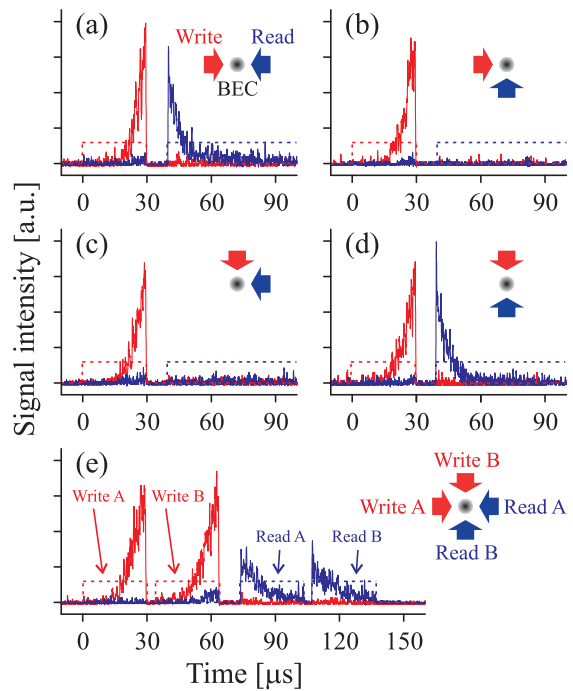}}
\end{center}
	\caption{(Color online). Demonstration of two-mode anti-Stokes and Stokes SRS. (a)-(d) Bragg-selective readout of the Stokes photons with four different write-read configurations (shown in the inset of each panel). (e) Superradiant writing/reading of two independent coherence gratings. The light intensity was fixed to $I_{\rm W} = 41$ mW/cm$^2$ for the write and $I_{\rm R} = 510$ mW/cm$^2$ for the read beam ($\gamma_{\rm S} / \gamma_{\rm AS} \sim 1$). Timing of the light beams is depicted by the dashed lines.}
	\label{Fig.3}
\end{figure}

Figures \ref{Fig.2}(c) and (d) show the momentum distribution of atoms after the write process and the successive read process, respectively. In the present case, the Stokes SRS is stimulated by the preformed grating, so that the system is closed within the three momentum states $|g \rangle$ and $|\pm \rangle$ as observed in (c). Consequently, all the recoiling atoms are pumped back into the initial state $|g \rangle$ and no higher-order momentum state appears as long as $n \ll N$. By contrast, in the case of $n \sim N$ where the BEC is significantly depleted, we have observed the onset of independent Stokes SRS, producing second-order momentum side modes $|1 \rangle \otimes |\vec{p} = \pm 2 \hbar \vec{k}_{\rm E} \rangle$ during the read process.

Next, we tried to verify the Bragg selectivity in the read process by changing the configurations of the write and the read beam. Figures \ref{Fig.3}(a)-(d) present the experimental results. As predicted by the Bragg condition, the Stokes photons could be retrieved only when the write-read beam pair was counter-propagating. Furthermore, by applying four laser beams sequentially, multiplexing of the write-read process could be realized as shown in Fig. \ref{Fig.3}(e). We observed the similar signals for any other write-read order and the signal magnitude for the write A (B) and the read A (B) was clearly correlated. 

Theoretically, the number of stored modes can be increased further by applying additional write-read beam pairs in $xy$ plane. However, it was difficult to achieve in the present setup due to the limitation of the optical access to BECs. Thus we only note the estimation of the maximum storage number $\alpha$ in the following simple model. When the read beam is tilted in $xy$ plane by an angle $\theta$ against the write-beam axis, the phase-matching direction of the read process is also tilted by $\theta$ against $z$ axis. If $\theta$ is larger than the diffraction angle of the end-fire modes $\theta_{\rm d}$, Stokes SRS will be strongly suppressed \cite{Black}. In other words, two gratings can be treated independently when the angle between their wave vectors is larger than $2 \theta_{\rm d}$. Therefore, the occupation angle per write-read beam pair is $\theta_{\rm m} = 2 \theta_{\rm d}$, and $\alpha = 2 \pi / \theta_{\rm m} \approx 55$ in the present condition.

\begin{figure}[tttt]
\begin{center}
    \scalebox{0.8}{\includegraphics{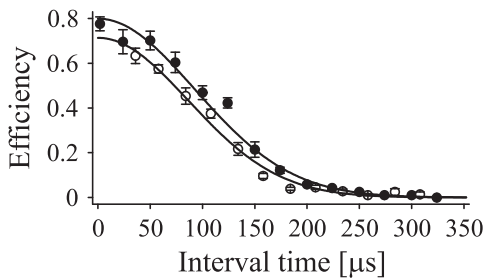}}
\end{center}
	\caption{Conversion efficiency versus interval time between the write and the read beam. Filled (open) circles represent the data for the single-mode (two-mode) storage of the grating.}
	\label{Fig.4}
\end{figure}

Finally, in Fig. \ref{Fig.4}, we plotted the conversion efficiency (the ratio of the anti-Stokes and the Stokes pulse area) versus the interval time between the write and the read beam. We fitted the data with a Gaussian function $P_0 \exp(-t^2 / \tau^2)$ and obtained $P_0 = 0.80(2)$, $\tau = 131(3)$ $\mu$s for the single-mode conversion and $P_0 = 0.71(2)$, $\tau = 122(4)$ $\mu$s for the two-mode conversion. The decrease of the efficiency for the later case might be due to the incoherent loss of the BEC in the state $|g \rangle$, which gives rise to decrease the contrast of all coherence gratings. A possible source of the lifetime degradation is the inhomogeneity of the light beams, which induces a position-dependent light shift and thus causes inhomogeneous phase evolution. The lifetime itself may be limited by the mean-field expansion of the BEC in free space \cite{Hagley}, but can be improved to $\sim 1$ ms by loading the atoms in an optical dipole trap, wherein the BEC and the recoiling atoms are both suspended by the same potential \cite{footnote}. Another way to achieve longer coherence time might be using ultracold fermions, in which the interatomic collisions are prohibited \cite{Roati}.

In summary, we have demonstrated superradiant writing and reading of two independent coherence gratings in an elongated Bose-Einstein condensate using angular multiplexing \cite{Heanue}. The present experiment was carried out in a range of the macroscopic photon number $n > 10^4$, which was too large to observe the quantum correlation between the anti-Stokes and the Stokes photons. However, this scheme would be readily extended to the quantum regime $n \sim 1$ by simply decreasing the write-beam intensity below the threshold \cite{Inouye1, Schneble1, Yoshikawa2}. Then, in combination with ``DLCZ'' scheme \cite{Duan1}, multiplexed generation, storage, and retrieval of quantum states in a single atomic cloud will be achieved. The observed high conversion efficiency of $> 70$\% and the long storage time of $> 120$ $\mu$s are due to the large optical density and the small Doppler width of BECs, that are the peerless advantage compared with the thermal atoms \cite{van der Wal, Chaneliere, Black, Yoshikawa2}. Furthermore, in BECs, it is possible to manipulate the stored momentum states via optically induced Bragg diffraction \cite{Hagley, Kozuma}. This ability facilitates the practical realization of quantum atom optics in an ensemble of atoms. For example, if we prepare two single-atom excitations (gratings) and mix them with a $\pi/2$-Bragg pulse, the atomic version of the two-photon Hong-Ou-Mandel interference would be realized \cite{Hong}.

\begin{acknowledgments}
We gratefully acknowledge A. Musha for assistance with the experiments. This work was supported by the Grants in Aid for Scientific Research from MEXT.
\end{acknowledgments}



\begin{thebibliography}{99}

\bibitem{Duan1}
L.-M. Duan {\it et al}., Nature (London) {\bf 414}, 413 (2001).

\bibitem{Lukin}
M. D. Lukin, P. Hemmer, and M. O. Scully, Adv. At. Mol. Opt. Phys. {\bf 42}, 347 (2000).

\bibitem{Chou1}
C. W. Chou {\it et al}., Phys. Rev. Lett. {\bf 92}, 213601 (2004).

\bibitem{Chaneliere}
T. Chaneli\`ere {\it et al}., Nature (London) {\bf 438} 833 (2005).

\bibitem{Black}
A. T. Black, J. K. Thompson, and V. Vuleti\'c, Phys. Rev. Lett. {\bf 95}, 133601 (2005).

\bibitem{van der Wal}
C. H. Van der Wal {\it et al}., Science {\bf 301}, 196 (2003).

\bibitem{Kuzmich}
A. Kuzmich {\it et al}., Nature(London) {\bf 423}, 731 (2003).

\bibitem{Balic}
V. Bali\'c {\it et al}., Phys. Rev. Lett. {\bf 94}, 183601 (2005).

\bibitem{Matsukevich}
D. N. Matsukevich and A. Kuzmich, Science {\bf 306}, 663 (2004).

\bibitem{Chou2}
C. W. Chou {\it et al}., Nature {\bf 438}, 828 (2005).

\bibitem{Duan2}
L.-M. Duan, J. I. Cirac, and P. Zoller, Phys. Rev. A {\bf 66}, 023818 (2002).

\bibitem{Horak}
P. Horak {\it et al}., Phys. Rev. A {\bf 67}, 043806 (2003); A. Haase, B. Hessmo, and J. Schmiedmayer, Opt. Lett. {\bf 31}, 268 (2006).

\bibitem{Gross}
M. Gross and S. Haroche, Phys. Rep. {\bf 93}, 301 (1982).

\bibitem{Inouye1}
S. Inouye {\it et al}., Science {\bf 285}, 571 (1999).

\bibitem{Schneble1}
D. Schneble {\it et al}., Phys. Rev. A {\bf 69}, 041601(R) (2004).

\bibitem{Yoshikawa1}
Y. Yoshikawa {\it et al}., Phys. Rev. A {\bf 69}, 041603(R) (2004).

\bibitem{Yoshikawa2}
Y. Yoshikawa, Y. Torii, and T. Kuga, Phys. Rev. Lett. {\bf 94}, 083602 (2005).

\bibitem{Heanue}
J. F. Heanue, M. C. Bashaw, and L. Hesselink, Science {\bf 265}, 749 (1994).

\bibitem{Dicke}
R. H. Dicke, Phys. Rev. {\bf 93}, 99 (1954).

\bibitem{Ketterle}
W. Ketterle and S. Inouye, Phys. Rev. Lett. {\bf 86}, 4203 (2001).

\bibitem{Slama}
S. Slama {\it et al}., Phys. Rev. Lett. {\bf 98}, 053603 (2007).

\bibitem{Roati}
G. Roati {\it et al}., Phys. Rev. Lett. {\bf 92}, 230402 (2004).

\bibitem{footnote}
In the case of superradiant Raman scattering, the Zeeman shift of the original and the recoiling states are different. Therefore, the inhomogeneous field of the magnetic trap drastically accelerates the decoherence of atoms and should be turned off during experiments.

\bibitem{Inouye2}
S. Inouye {\it et al}., Phys. Rev. Lett. {\bf 85}, 4225 (2000).

\bibitem{Schneble2}
D. Schneble {\it et al}., Science {\bf 300}, 475 (2003).

\bibitem{Zobay}
O. Zobay and G. M. Nikolopoulos, Phys. Rev. A {\bf 72}, 041604(R) (2005).

\bibitem{Hagley}
E. W. Hagley {\it et al}., Phys. Rev. Lett. {\bf 83}, 3112 (1999).

\bibitem{Kozuma}
M. Kozuma {\it et al}., Phys. Rev. Lett. {\bf 82}, 871 (1999).

\bibitem{Hong}
C. K. Hong, Z. Y.  Ou, and L. Mandel, Phys. Rev. Lett. {\bf 59}, 2044 (1987).

\end{thebibliography}
\end{document}